\documentstyle[12pt]{article}  
\date{ }

\title{Strongly gravitating empty spaces}
\author{J. S\l adkowski \dag \\  Institute of Physics, 
University of Silesia, \\ Uniwersytecka 4, Pl 40007 Katowice,\\  Poland}

\begin{document}
\baselineskip9mm
\maketitle
\begin{abstract}
\baselineskip7mm
We use various results concerning isometry groups of Riemannian and 
pseudo-Riemannian manifolds to prove that there are spaces on which 
differential structure can act as a source of gravitational force 
(Brans conjecture). The 
result is important for the analysis of the possible  physical meaning 
of differential calculus. Possible astrophysical consequences are discussed. 

\end{abstract}

\vspace{40mm}
\dag e-mail: sladk@uranos.cto.us.edu.pl

\newtheorem{theorem}{Theorem}
\newpage
\section{Introduction}

\ \ \ The choice of mathematical model for spacetime has important 
physical significance. B. Riemann suggested that the geometry of space 
may be more than just a mathematical tool defining a stage for 
physical phenomena, and may in fact have profound physical meaning 
in its own right [1]. With the advent of general relativity 
theoreticians started to think of the spacetime as a differential 
manifold. Since then various assumptions about 
the spacetime topology and 
geometry have been discussed in the literature [2]. Until recently, 
the choice of differential structure of the spacetime manifold 
has been assumed to be trivial because most topological spaces used 
for modeling spacetime have natural differential structures and these 
differential structures where (wrongly) thought to be unique. Therefore 
the counterintuitive discovery of exotic ${\bf R}^{4}$'s following 
from the work  of Freedman [3] and Donaldson [4] raised various 
discussions about the possible physical consequences of this discovery. 
Exotic ${\bf R}^{4}$'s are smooth 
($C^{\infty}$) four-manifolds which are homeomorphic to the 
Euclidean four-space ${\bf R}^{4}$ but not diffeomorphic to it. Exotic 
${\bf R}^{4}$'s are unique to dimension four, see [5-11] for details. 
Later we have realized that exotic (nonunique) smooth structures are 
abundant in dimension four. For example it is sufficient to remove 
one point from a given four-manifold to obtain a manifold with exotic 
differential structures [11]; every manifold of the form $M\times 
{\bf R}$, $M$ being compact 3-manifold has infinitely many inequivalent 
differential structures. Such manifolds play important role in 
theoretical physics and astrophysics. 
Therefore the physical meaning of exotic smoothness 
must be thoroughly investigated. This is not an easy task: we only know 
few complicated coordinate descriptions [12] and most mathematicians believe 
that there is no finite atlas on an exotic ${\bf R}^{4}$. 
To our knowledge, only few physical 
examples have been discussed in the literature [2,6,7,13,14]. In this paper 
we would like to discuss some peculiarities that may happen while studying 
the theory of gravity on some exotic ${\bf R}^{4}$'s. 
The most important result is that on some 
topologically trivial spaces there exist only "complicated" solutions 
solutions to the Einstein equations. By this we mean that there may be 
no stationary cosmological model solutions 
and/or that empty space can gravitate. Such solutions are 
counterintuitive but we are aware of no physical principle that would require 
rejection of such spacetimes (besides common sense?). 

\section{General relativity on exotic ${\bf R}^{4}$'s with few symmetries}

\ \ \ As it was written in the previous section, exotic ${\bf R}^{4}$'s 
are defined as four-manifolds that are homeomorphic to the fourdimensional 
Euclidean space ${\bf R}^{4}$ but not diffeomorphic to it. There are 
infinitely many of such manifolds (at least a two parameter family of them) 
[5]. Note that exotic differential structures do not change the definition 
of the derivative. The essential difference is that the rings of real 
differentiable functions are different on nondiffeomorphic manifolds. In 
the case of exotic ${\bf R}^{4}$'s this means that there are some 
continuous functions  ${\bf R}^{4} \mapsto {\bf R}$ that are smooth 
on one exotic ${\bf R}^{4}$ and only continuous on another and vice versa [9]. 
To proceed we will recall several definitions. We will call a diffeomorphism 
$\phi \ :\ M \mapsto M$, where $M$ is a (pseudo-)Riemannian manifold with 
metric 
tensor $g$ an isometry if and only if it preserve $g$, $\phi ^{*} g=g$ [15]. 
Such mappings form a group called the isometry group. We say that a smooth 
manifolds has few symmetries provided that for every choice of differentiable 
metric tensor, the isometry group is finite. Recently, L. R. Taylor managed to 
construct examples of exotic ${\bf R}^{4}$'s with few symmetries [16]. 
Among these there are examples with nontrivial but still finite isometry 
groups. Taylor's result, although concerning Riemannian 
structures, has profound 
consequences  for the analysis of the possible role of differential structures 
in physics where Lorentz manifolds are commonly used. To show this let us 
define a (non-)proper actions of a group on manifolds as follows. Let $G $ be a 
locally compact topological group acting on a metric space $X$. We say that 
$G$  acts properly on $X$ if and only if for all compact subsets $Y 
\subset X$, the set $\{ g\in G:gY\cap Y\not= \emptyset \}$ is also
compact. 
Restating this we say that $G$ acts nonproperly on $X$ if and only if there 
exist sequences $x_{n}\rightarrow  x$ in $X$ and $g_{n} \rightarrow \infty $ 
in $G$, such that $g_{n}x_{n}$ converges in $X$. Here $g_{n} \rightarrow 
\infty $ means that the sequence $g_{n}$ has no convergent subsequence in the 
compact open topology on the set of all isometries [15 p. 202]. 
Note that for many 
manifolds a proper $G$ action is topologically impossible and on the other 
hand a nonproper $G$ action on a Lorentz (or pseudo-Riemannian) manifolds is 
for all but a few groups also impossible [17, 18]. Our discussion would 
strongly depend on the later fact and on the theorems proved by N. Kowalsky 
[18]. First of all let us quote [18]: 
\begin{theorem}
Let G be Lie transformation group of a differentiable manifold X. If G acts 
properly on X, then G preserves a Riemannian metric on X. The converse is true 
if G is closed in Diff(X). 
\end{theorem}
As a special case we have:
\begin{theorem}
Let G and X be as above, and in addition assume G connected. If G acts 
properly on X preserving a time-orientable Lorentz metric, then G preserves 
a Riemannian metric and an everywhere nonzero vector field on X.
\end{theorem}
If we combine these theorems with the Taylor's results we immediately get:
\begin{theorem}
Let G be a Lie transformation group acting properly on an exotic 
${\bf R^{4}}$ with few symmetries and preserving a time-orientable 
Lorentz metric. Then G is finite. 
\end{theorem}
Further, due to N. Kowalsky, we also have [18]:
\begin{theorem}
Let G be a connected noncompact simple Lie group with finite center. Assume 
that G is not locally isomorphic to SO(n, 1) or SO(n,2). If G acts 
nontrivially on a manifold X preserving a Lorentz metric, then G 
actually acts properly on X.
\end{theorem} 
and 
\begin{theorem}
If G acts nonproperly and nontrivially on X, then G must be locally isomorphic 
to SO(n,1) or SO(n,2) for some n.
\end{theorem}
The general nonproper actions of Lie groups locally isomorphic to SO(n,1) 
or SO(n,2) would be discussed in ref. [19]. In many cases it is possible to 
describe the cover $\tilde X$ up to Lorentz isometry. \\ 

Now, suppose we are given an exotic ${\bf R}_{\theta}^{4}$ 
with few symmetries. Given any boundary conditions, we can try to solve the  
Einstein equations on ${\bf R}_{\theta}^{4}$. Suppose we have found some 
solution to the Einstein equations on ${\bf R}_{\theta}^{4}$. Whatever the 
boundary conditions be we would face one of the two following situations. 
\begin{itemize}
\item The isometry group G of the solution  acts properly on 
${\bf R}_{\theta}^{4}$. Then according to Theorem 3 G is finite. There is no 
nontrivial Killing vector field and the solution cannot be stationary [19]. 
The gravitation is quite "complicated" and even empty spaces do evolve. Note 
that this conclusion is valid for any open subspace of ${\bf R}_{\theta}^{4}$. 
This means that this phenomenon cannot be localized on such spacetimes. 
\item The isometry group G of the solution  acts nonproperly on 
${\bf R}_{\theta}^{4}$. Then G is locally isomorphic to SO(n,1) or SO(n,2 ). 
But the nonproper action of G on ${\bf R}_{\theta}^{4}$ means that there are 
points infinitely close together in ${\bf R}_{\theta}^{4}$ ($x_{n} \rightarrow 
x$) such that arbitrary large different isometries ($g_{n} \rightarrow 
\infty $) in G maps them  
into infinitely close points in ${\bf R}_{\theta}^{4}$ 
($g_{n}x_{n} \rightarrow y \in {\bf R}_{\theta}^{4}$). There must exists quite 
strong gravity centers to force such convergence (even in empty spacetimes). 
Such spacetimes are unlikely to be stationary.
\end{itemize}

We see that in both cases Einstein gravity is quite 
nontrivial even in the absence 
of matter. Let us recall that if a spacetime has a Killing vector field 
$\zeta ^{a}$, then every covering manifolds admit appropriate Killing vector 
field $\zeta ^{'a}$ such that it is projected onto $\zeta ^{a}$ by the 
differential of the covering map. This means that discussed above 
properties are 
"projected" on any space that has exotic ${\bf R}^{4}$ with few symmetries as 
a covering manifold eg quotient manifolds obtained by a smooth action 
of some finite group. Note that we have proven a weaker form of the 
Brans conjecture [7]: 
there are four-manifolds (spacetimes) on which differential 
structures can act as a source of gravitational force just as ordinary 
matter does.

\section{Conclusions}
The existence of topologically trivial spacetimes that admit only 
"nontrivial" solutions to the Einstein equations is very surprising. Such 
phenomenon might be also possible for other four-manifolds admitting exotic 
differential structures enumerated in the Introduction. The first reaction is 
to reject them as being unphysical mathematical curiosities. But this 
conclusion might be erroneous [6-8, 13]. Besides the arguments put forward by 
Brans [6-8] and Asselmeyer [13] we would like to add the following. Suppose that 
spacetime is only a secondary entity emerging  as a result of interactions 
between physical (matter) fields. If we use the A. Connes' noncommutative 
geometry formalism do describe Nature then Dirac operators and their spectra 
define the spacetime structure [22, 23]. There are known 
examples of differential structures that are 
distinguished by spectra of the Dirac operators [23-25]. This 
suggest that fundamental interactions of matter are "responsible" for the 
selection of the differential  structure and might not have "chosen" the 
simplest 
structure of the spacetime manifold. Although it is unlikely that the spectrum 
of the Dirac operator alone would allow to distinguish the differential 
structure [26] in a general case, it seems to be 
reasonable to conjecture that physically equivalent
 spacetimes must be isospectral  [2] and we might hope that this would 
solve the problem with  the plethora of exotic 
differential structures. If Nature 
have not used exotic smoothness we physicists should find out why only one 
of the existing differential structures has been preferred. Does it mean that 
the differential calculus, although very powerful, is not  necessary 
(or sufficient) for the description of the laws of physics? It might not be 
easy to find any answer to these questions. \\ 

\ \ \ Let us conclude  by saying that if exotic smoothness has anything to do 
with the physical world it may be a source/ explanation of various 
astrophysical and cosmological phenomena. Dark matter and vacuum energy 
substitutes and attracting centers are the most obvious among them 
[27-29]. "Exoticness" of the spacetime might be responsible for the recently 
discovered anomalies in the large redshift supernovae properties. 
The process of "elimination" of exotic differential structures 
might also result in the emergence time [30, 31] or spacetime signature. \\

\ \ \ {\bf Acknowledgments}. The author would like to 
thank K. Ko\l odziej and I. Bednarek 
for stimulating and helpful discussions. 

\newpage
\subsection*{\ \ References}

\newcounter{bban}

\begin{list}
{[\arabic{bban}.]}{\usecounter{bban}\setlength{\rightmargin}
{\leftmargin}}
\item Riemann B, \" Uber die Hypothesen, welche der Geometrie zugrunde 
liegen (1854). 
\item S\l adkowski J, hep-th/9610093.
\item Freedman M H, J. Diff. Geom. 17 (1982) 357.
\item Donaldson S K, J. Diff. Geom 18 (1983) 279.
\item Gompf R, J. Diff. Geom. 37 (1993) 199.
\item Brans C and Randall D, Gen. Rel. and Grav. 25 (1993) 205.
\item Brans C, Class. Quantum. Grav. 11 (1994) 1785.
\item Brans C, J. Math. Phys. 35 (1994) 5494.
\item S\l adkowski J, Acta Phys. Pol. B27 (1996) 1649.
\item Freedman M H and Taylor L R, J. Diff. Geom. 24 (1986) 69.
\item Bizaca Z and Etnyre J, Topology 37 (1998) 461.
\item Bizaca Z, J. Diff. Geom. 39 (1994) 491.
\item Asselmeyer T, Class. Quantum Grav. 14 (1997) 749.
\item S\l adkowski J, Int. J. Theor. Phys. 33 (1994) 2381.
\item Helgason S, Differential Geometry, Lie Groups and Symmetric 
spaces, Academic Press, 1978.
\item Taylor R L, math.QT/9807143.
\item Szaro J P, Am. J. Math. 120 (1998) 129.
\item Kowalsky N, Ann. Math. 144 (1997) 611.
\item Kowalsky N, Actions of SO(n,1) and SO(n,2) on Lorentz 
manifolds, in preparation.
\item Wald R, General Relativity, The Univ. of Chicago Press, Chicago (1984).
\item Connes A, Noncommutative Geometry, Academic Press London 1994. 
\item S\l adkowski J, Acta Phys. Pol. B 25 (1994) 1255. 
\item Donnelly H, Bull. London Math. Soc. 7, 147 (1975).
\item Kreck M and Stolz S, Ann. Math. 127 (1988) 373.
\item Stolz S, Inv. Math. 94 (1988) 147.
\item Fintushel R and Stern R J, math.SG/9811019. 
\item Bednarek I and Ma\' nka R, Int. J. Mod. Phys. D 7 (1998) 225.
\item Ma\' nka R and Bednarek I, Astron. Nachr. 312 (1991) 11.
\item Ma\' nka R and Bednarek I, Astrophys. and Space Science 176 
(1991) 325.
\item Isham C and Butterfield J, in The Arguments of Time, J. Butterfield 
(ed.), Oxford Univ. Press, Oxford 1999.
\item Heller M and Sasin W, Phys. Lett A 250 (1998) 48.
\end{list}

\end{document}